\documentclass[twoside]{article}

\usepackage{graphicx}
\usepackage{cmpj2eMod}
\title[Gauge field theory approach to spin transport]%
	{Gauge field theory approach to spin transport in a 2D electron gas}

\author[B. Berche, N. Bol\'\i var, A. L\'opez and E. Medina] 
	{B. Berche\refaddr{label1,label2}, 
	N. Bol\'\i var\refaddr{label3}, 
	A. L\'opez\refaddr{label2} 
	and E. Medina\refaddr{label1,label2,label3}}
\addresses{
	\addr{label1} \ Statistical Physics Group, 
	P2M Department, Institut Jean Lamour,\\ 
        UMR CNRS 7198, BP 70239\\
        F-54506 Vand\oe uvre les Nancy Cedex, France\\	
	\addr{label2} \ Centro de F\'\i sica,\\	
	Instituto Venezolano de Investigaciones Cient\'\i ficas,\\
	Apartado 21827, Caracas 1020A, Venezuela\\
	\addr{label3}\ Departamento de F\'\i sica, 
	Universidad Central de Venezuela,\\ 
	Caracas, Venezuela}

\begin{document}

\maketitle
\begin{abstract}
We discuss the Pauli Hamiltonian including the spin-orbit interaction within 
an $U(1)\times{SU(2)}$ gauge theory 
interpretation, where the gauge symmetry appears to be broken. 
This interpretation offers new insight into the problem of 
spin currents in the condensed matter environment, and can be extended
to Rashba and Dresselhaus spin-orbit interactions. 
We present a few outcomes of the present formulation: i) it automatically
leads to zero spin conductivity, in contrast to 
predictions of Gauge symmetric treatments, ii) a topological quantization
condition leading to voltage quantization follows, and iii) spin
interferometers can be conceived in which, starting from a arbitrary incoming
unpolarized spinor, it is always possible to construct
a perfect spin filtering condition.


 
\end{abstract}

\def\be{\begin{equation}}
\def\ee{\end{equation}}
\def\beq{\begin{eqnarray}}
\def\eeq{\end{eqnarray}}
\def\bpm{\begin{pmatrix}}
\def\epm{\end{pmatrix}}
\newcommand{\<}{\langle}
\renewcommand{\>}{\rangle}
\newcommand{\x}{{\boldsymbol x}}
\newcommand{\y}{{\boldsymbol y}}
\renewcommand{\k}{{\boldsymbol k}}
\newcommand{\q}{{\boldsymbol q}}
\def\bmu{{\boldsymbol \mu}}
\def\bsigma{{\boldsymbol\sigma}}
\def\btau{{\boldsymbol\tau}}
\def\boldsymbol#1{\mbox{\boldmath $#1$}}
\newcommand{\bfa}{{\cal W}\!\!\!\!\!\!{\cal W}}
\def\nbOne{\hbox{{\bf 1$\!\!$l}}}
\def\nbM{\hbox{{\bf I\hskip-.8mm M}}_{2\times 2}}

\section{Introduction}
\label{Intro}

A reformulation of the spin-orbit (SO) coupling Hamiltonian in terms of 
non-Abelian gauge fields~\cite{ryder} was explicitly given in 
ref. \cite{Rebei,Jin,Leurs,Medina} where the SO interaction is presented 
as a  $SU(2)\times U(1)$ gauge theory. As the Yang-Mills gauge theory is 
well understood and is the underpinning of well established theory, 
enormous insight can be brought upon new problems. Such gauge point of 
view, in more general terms, has been known for some 
time~\cite{Goldhaber,Mineev,Frohlich,FrohlichCMP}. 
This formulation is very revealing, 
since the consistent gauge structure of the theory becomes obvious and the 
physics of spin currents, persistent currents and color 
diamagnetism~\cite{Tokatly} can be understood in a manner analogous to the 
well known $U(1)$ gauge theories. A consistent  $SU(2)\times U(1)$ gauge 
approach was presented in reference \cite{Leurs,Medina} where it was found 
that for the Pauli type Hamiltonians (including Rashba and two-dimensional 
reductions of the Dresselhaus Hamiltonian), Gauge Symmetry Breaking (GSB) 
is necessarily built into the theory and leads to vanishing of the spin 
conductivity in constant electric fields~\cite{Medina}. In addition, the 
Yang Mills interpretation of the Rashba and Dresselhaus SO interactions 
renders the associated gauge fields real, with topological consequences 
analogous to the Aharonov Casher effect~\cite{Leurs,Medina}. 

The Rashba and Dresselhaus SO interactions arise in materials which lack 
either structural or bulk inversion symmetry, 
respectively~\cite{Rashba60,Dresselhaus,winkler}. These two kinds of 
interactions have recently been given a great deal of attention due to 
their potential role in the generation and manipulation of spin polarized 
currents, spin filters~\cite{Nitta}, spin accumulation~\cite{SarmaReview}, 
and spin optics~\cite{BalseiroUsaj}.

\section{Spin-orbit interaction in semiconductors
	\label{Sec:Intro}}
Spin-orbit  interaction may be defined as the interaction of the spin
magnetic moment of the electron, ${\vec\bmu}=-g\frac{e}{2m}{\vec {\bf s}}=
-\frac{e\hbar}{2m}{\vec\bsigma}$ with the magnetic field produced by all
external moving charges in the electron rest frame. Here, 
 ${\vec {\bf s}}=\frac 12\hbar{\vec \bsigma}$ is the spin of the electron with
${\vec \bsigma}$ the vector of Pauli matrices, 
$g\simeq 2$ the Land\'e factor, $m$ the electron rest mass and $-e$ the 
electron electric charge. Bold face symbols denote matrices in spin space.
In the rest frame of the external charges, the SO interaction, properly 
corrected to take into account Thomas precession, 
is usually described by the Pauli term, $\sim 
\frac 1{2m^2c^2}{\vec {\bf s}}\cdot({\vec p}\times{\vec \nabla}V)$, where
$V$ is the potential energy of the electron in the presence of 
the external charges.  
For a spherically symmetric potential, ${\vec \nabla}V= f(r){\vec r}$
and the SO interaction is proportional to 
${\vec s}\cdot{\vec L}$,
hence its name.

In a semi-conductor, the electron rest mass and the Land\'e factor must
be replaced by their effective counterparts, and the spin-orbit interaction can
thus be considerably enhanced. Moreover, an appropriate basis in a crystalline
solid is given in terms of the Bloch wave functions, 
$\psi({\vec r})=u_{{\vec k}}e^{i{\vec k}{\vec r}}$, and the 
SO interaction must be calculated in this basis. This is generally a hard task 
and one usually uses phenomenological expressions compatible with the
crystal symmetries.
When there is bulk inversion asymmetry (BIA), we have the Dresselhaus
expression in $3d$ systems~\cite{EngelRashbaHalperin},
\begin{equation}
H_{D,3d}={\rm const.}\ \!k_x(k_y^2-k_z^2)\bsigma_x+{\rm c.p.}
\end{equation}
where ${\rm c.p.}$ stands for cyclic permutations. When the electrons are 
confined in two dimensions, the expectation value along the third
dimension should be considered, with $\langle k_z\rangle\simeq 0$,
$\langle k_z^2\rangle\simeq(\pi/a)^2$, $a$ being the typical confinment 
length in the $z-$direction. 
The Dresselhaus SO interaction thus takes the simple form
\begin{equation}
H_{D,2d}=\beta (k_x\bsigma_x-k_y\bsigma_y)
\end{equation}
where we neglect cubic terms in $k$.
If the confining potential is not symmetric i.e. in the case of space
inversion asymmetry (SIA), there is another term which directly follows from
the expression $\frac{\hbar^2}{4m^2c^2}{\vec \bsigma}\cdot({\vec k}\times 
{\vec \nabla} V)$ with only the $\langle\nabla_z V\rangle$ contribution.
This term is known as the Rashba SO interaction,
\begin{equation}
H_{R,2d}=\alpha (k_y\bsigma_x-k_x\bsigma_y).
\end{equation}
Note that the Rashba SO amplitude can be tuned experimentally using a gate
voltage, since the coefficient $\alpha$ is proportional to the electric
field. 
For more details on SO interactions in semi-conductors, see 
Refs.~\cite{winkler,EngelRashbaHalperin}.

\section{Non-Abelian gauge field theory approach\label{Sec:Gauge}}

Let us first consider the Pauli SO interaction. Rashba and Dresselhaus SO
interactions will be treated later.
Neglecting electron-electron interactions, we start by considering the 
Pauli Hamiltonian to order $v/c$, acting on two-component spinors,
\begin{eqnarray}
H&=&
\Bigl[\frac{({\vec p}-e{\vec A})^2}{2m}
+V-e\phi\Bigr]\nbOne_{2\times 2}  
-\Bigl[\frac{{\vec p\ \!}^4}{8m^3c^2}
+\frac{e\hbar^2}{8m^2c^2}\nabla\cdot{\vec E}\Bigr]
\nbOne_{2\times 2}\nonumber\\
&&+\frac{e\hbar}{2m}{\vec \bsigma}\cdot {\vec B}
-\frac{e\hbar{\vec \bsigma}\cdot ({\vec p}-e{\vec A})\times{\vec E}}{4m^2c^2}
\mbox{},\label{PauliFirst}
\end{eqnarray}
where the first term in
the first line corresponds to the usual Schr\"odinger equation including
the kinetic energy with a mininal coupling  to the electromagnetic field,  
the substrate potential denoted by $V$, that can be assumed periodic, and 
a scalar potential contribution.  
The second term in the first line describes the 
first relativistic correction to the kinetic energy and the Darwin term,
where ${\vec E}$ is the electric field and $c$ the speed of light. These
 first two terms are proportional to the $2\times 2$ identity matrix
in spin space. The second
line comprises explicitly spin-dependent terms, first the Zeeman
interaction where $\vec B$ is the magnetic field and $\vec \bsigma$ 
is the Pauli matrix vector and the second term is the spin-orbit interaction,
now written with the minimal coupling to the gauge vector. 
We have assumed a static potential so that the rotor of the electric field 
is absent and the spin-orbit interaction is limited to the term mentioned 
here.  
In the following, we absorb the spin-independent one-body interactions 
(second term of first line in Eq.~\ref{PauliFirst}) 
in the potential, and we rewrite
the non relativistic kinetic energy plus the spin-orbit interaction as
\beq
{\rm K.E.}+{\rm S.O.}&=&\frac 1{2m}\left[
({\vec p}-e{\vec A})^2\nbOne_{2\times 2}
-2({\vec p}-e{\vec A})\nbOne_{2\times 2}
\frac{-e\hbar}{4mc^2}{\vec \bsigma}\times{\vec E}\right]
\nonumber\\
&=&\frac 1{2m}\Bigl[
({\vec p}-e{\vec A})^2\nbOne_{2\times 2}
-\frac{-e\hbar}{4mc^2}{\vec \bsigma}\times{\vec E}\Bigr]^2
-\frac{e^2\hbar^2}{32m^3c^4}|{\vec \bsigma}\times{\vec E}|^2.
\eeq
This suggests an $SU(2)\times U(1)$ form described by non-Abelian 
$W^{\mu a}$ and ordinary Abelian gauge fields 
$A^\mu=(A^0,A^i)=(\phi/c,{\vec A})$ with ${\vec E}=-{\vec \nabla}\phi
-\partial_t{\vec A}$ and ${\vec B}={\vec \nabla}\times{\vec A}$, 
and we can rewrite 
the Hamiltonian, following Jin, Li and Zhang~\cite{ZhangYangMills} as
\begin{eqnarray}
\label{HLambda}
H&=&\frac{1}{2m}\Bigl[ ({\vec p}-e{\vec A})\nbOne_{2\times 2}
-g{\vec W}^a\btau^a\Bigr]^2\nonumber\\
&&-\frac 1{8m}g^2{\vec W}^a{\vec W}^a\nbOne_{2\times 2}
-gcW^{0a}\btau^a+(V'-ecA^0)\nbOne_{2\times 2}
,\label{PauliSecond}
\end{eqnarray}
where the Zeeman interaction is written as the time component of the 
non-Abelian gauge field~\cite{ZhangYangMills} 
$-gcW^{0a}\btau^a=\frac{e\hbar}{2m}{\vec \bsigma}\cdot {\vec B}$, while
the space components of this $SU(2)$ connection are defined by 
$g{W}^{ia}\btau^a=-(e\hbar/2 mc^2)\varepsilon_{iaj}{ E}^j\btau^a$,
or explicitly,
\beq
g{\vec W}^1&=&\frac{e\hbar}{2mc^2}(E_z\vec {u_y}-E_y\vec {u_z}),
\\
g{\vec W}^2&=&\frac{e\hbar}{2mc^2}(-E_z\vec {u_x}+E_x\vec {u_z}),
\\
g{\vec W}^3&=&\frac{e\hbar}{2mc^2}(E_y\vec {u_x}-E_x\vec {u_y}).
\eeq
with $\vec {u_x}$, $\vec {u_y}$ and $\vec {u_z}$ unit vectors in the $x-$, 
$y-$ and
$z-$directions.
 The $2\times 2$ matrices 
$\btau^a$ are the symmetry generators for $SU(2)$ obeying the commutation 
relation $[\btau^a,\btau^b]=i\varepsilon_{abc}\btau^c$, $\varepsilon_{abc}$ 
being the totally antisymmetric  tensor. The coupling constant $g$ is fixed 
by the combination $gW^{\mu a}$ which has the dimensions 
of $\frac{e\hbar}{2mc^2}E$, and we choose
$g=\hbar$. 
The relation between the spin operator and the corresponding generators 
is $\hbar\btau^a={\bf s}^a$ and the spin is $s=1/2$.
The superscripts of the beginning of the Latin alphabet, $a,b,c,\dots$ refer
to the internal spin degrees of freedom for which we use the convention 
of summation when they are repeated, while Greek  indices $\mu,\nu,\ \dots$ 
correspond to space-time components and run from $0$ to $3$, 
the time component corresponding to $0$ and the space components being also
denoted as Latin indices from the middle of the alphabet, $i,j,k$, \dots

This formulation differs from that of ref.~\cite{ZhangYangMills,Tokatly} most 
importantly in the term quadratic in ${\vec W}^a$ 
in equation~\ref{PauliSecond}.
The purpose of writing the second term here, as a function of the 
$SU(2)$ connection is to evidence {\it gauge symmetry breaking} (GSB) in this 
Hamiltonian. This observation has important consequences in the physical 
interpretation of the resulting Yang-Mills fields and is the reason why the 
Yang-Mills fields themselves are observable quantities, where as in a gauge 
symmetric theory they would be gauge dependent~\cite{weinberg,Hughes}.

In order to generate the Noether currents in a canonical fashion, one must formulate the appropriate Lagrangian for the corresponding equations of motion. The non-relativistic Lagrangian density we seek is (now
omitting identity matrices) 
\begin{eqnarray}
{\mathcal L}&=& 
\frac{i\hbar}{2}
	\bigl(\Psi^{\dagger}\dot{\Psi}-\dot{\Psi}^{\dagger}\Psi\bigr)
-\frac{1}{2m}
	\bigl( {-i\hbar\vec {\mathcal D}}\Psi\bigr)^{\dagger}
	\bigl({-i\hbar\vec {\mathcal D}}\Psi\bigr)
	\nonumber\\
&&-
	\Psi^{\dagger}
	\Bigl(\frac{-g^2}{8m}{\vec W}^b{\vec W}^{b}
	+gc{W}^{0a}\btau^a+ec{A}^0\Bigr)
	\Psi
	\nonumber\\
&&-\frac{e^2}{4m}F_{\mu\nu}F^{\mu\nu}
-\frac{g^2}{4m}{G}^a_{\mu\nu}{ G}^{\mu\nu a},
\end{eqnarray}
where $\Psi$ is a Pauli spinor 
$\Psi=\Bigl(\begin{array}{c}\psi_\uparrow \\ 
\psi_\downarrow\end{array}\Bigr)$,
${G}_{\mu\nu}^a =\partial_\mu{W}_\nu^a
-\partial_\nu{W}_\mu^a
 -\epsilon^{abc}{W}_\nu^b{W}_\mu^c$ and ${F}_{\mu\nu} 
=\partial_\mu A_\nu-\partial_\nu A_\mu$ are the $SU(2)$ and $U(1)$ field 
tensors respectively. The new term $-\frac{g^2}{8m}\Psi^\dagger{\vec W}^b
{\vec W}^{b}\Psi$ is 
due to gauge symmetry breaking. The covariant derivative is then of the 
form 
$-i\hbar{\vec {\mathcal D}}=-i\hbar{\vec\nabla}-e{\vec A}-g{\vec W}^a\btau^a$.
 This form of the covariant derivative determines the well known $U(1)$ 
coupling constant $e/\hbar$ and the $SU(2)$ coupling constant for this theory 
is $g/\hbar$. 
A gauge transformation of the Lagrangian density would leave it unchanged
up to a divergence at the condition that the Coulomb gauge for the $SU(2)$
connection is satisfied, ${\vec\nabla}\cdot{\vec W}^a=0$.

The Hamiltonian in Eq.~\ref{PauliSecond} is derived from the 
corresponding Lagrange equations for the matter fields $\Psi$. 
The equations of motion of the Yang-Mills fields in the presence of currents
(generalized Maxwell equations) follow from Euler-Lagrange equations with
respect to variations of the gauge fields, 
 \begin{eqnarray}
 \partial_\mu\frac{\partial{\mathcal L}}
{\partial(\partial_\mu {W}^a_\nu)}
=\frac{\partial {\mathcal L}}{\partial{W}^a_\nu}
.
 \end{eqnarray}
The l.h.s. of this equation gives
$ \partial_\mu\frac{\partial{\mathcal L}}
{\partial(\partial_\mu {W}^a_\nu)}=
-\partial_\mu{G}^{\mu\nu a}$, and since the non-Abelian field tensor
${G}^{\mu\nu a}$ is antisymmetric with respect to space-time indices,
it is natural to introduce a conserved current
${\mathcal J}^{\nu a}=\frac{\partial {\mathcal L}}{\partial{W}^a_\nu}$.
The  conserved current 
${\mathcal J}^{\nu a}$~\cite{weinberg} is the full spin current 
carried both by matter and radiation i.e. $\vec{\mathcal J}^a
=\vec{\mathcal J}^a_{\rm Matter}+\vec{\mathcal J}^a_{\rm Radiation}$. This full current is an observable
physical quantity, since the gauge is fixed in the present case (in
contradistinction with ordinary $SU(2)$ gauge theory where it is 
gauge-dependent).
The spatial (spin current) 
and time (magnetization) components of the current density then follow as
\begin{eqnarray}
\label{totalspatialcurrent}
\vec{\mathcal J}^{a}&=&
\frac{g}{2m}\Bigl[ \left 
(\btau^a\Psi\right)^{\dagger}\bigl(-i\hbar \vec{\mathcal D}\Psi\bigr)
+\bigl(-i\hbar \vec{\mathcal D}\Psi\bigr)^{\dagger}(\btau^a\Psi) \Bigr]
\nonumber\\
&&\qquad+\Psi^{\dagger} \Bigl( \frac{g^2}{4m}\vec {W}^{a}\Bigr)\Psi\nonumber\\
&&\qquad\quad+\frac{g^2}{m}\varepsilon_{abc}{W}_\nu^b
({G}^{\nu xc}\vec u_x+{G}^{\nu yc}\vec u_y+{G}^{\nu zc}\vec u_z)
,\label{eqCurrent}
\end{eqnarray}
and the spin polarization
\begin{eqnarray}
\label{magnetization}
{\cal J}^{0a}=\Psi^{\dagger}g\btau^a\Psi+\frac{g^2}{m}
\varepsilon_{abc}{W}^b_j{G}^{j0c}.
\end{eqnarray}
Three terms can be distinguished in the spin current.
\begin{description}
\item i) The first term has the 
canonical form expected for the material current namely 
\be\vec{\mathcal J}_{\rm Matter}^{a}
=(g/2)\Psi^{\dagger}(\btau^a \vec v+\vec v\btau^a)\Psi,\ee 
where 
$v^i=(1/i\hbar)[r^i,H]$. 
\item ii) The second term comes from the gauge symmetry breaking contribution.
It depends on both matter and radiation.
\item iii) Finally, 
the third term is the canonical radiative contribution originating from the 
derivative with respect to the gauge potential of the non-Abelian 
contribution of the field tensor ${G}^{a}_{\mu\nu}$. Note that this last
term would not arise in an Abelian theory (e.g. in $U(1)$, since
the photon does not carry any electric charge).
\end{description} 
The first and last terms were described in ref.~\cite{ZhangYangMills} 
as taken from an apparently gauge symmetric form. The magnetization term has 
both a material contribution (the first term) and a radiative contribution 
as both matter and radiation carry angular momentum. We emphasize that, 
the extent to which gauge symmetry is broken depends on the choice of the 
electric field. If only the $E_z\ne 0$, then one allows gauge transformations 
that leave $W_1^2=-W_2^1$ invariant. This is analogous to the remnant $Z_2$ 
group after $U(1)$ GSB in superconductors and a similar situation in the 
electro-weak GSB mechanism~\cite{weinberg}.

Since the gauge theory considered here is non-Abelian, in the most general case
 the corresponding 
current is not gauge invariant (as it would be in the $U(1)$ case i.e. 
electromagnetism) and thus a function that is not 
measurable~\cite{Comment}. An important 
property in the SO case is that the gauge field is determined by the 
physical electric field, and thus the gauge is fixed and the spin current 
becomes properly defined. The third term in Eq.~\ref{eqCurrent}, 
which has the structure 
of a gauge symmetry breaking term, would also by itself fix the gauge and 
it has further consequences to be discussed below.

Rewriting the last term in Eq.\ref{totalspatialcurrent} in terms of ordinary 
derivatives plus a gauge dependent term we obtain
\be
-\frac{i\hbar g}{2m}\Bigl[ \left 
(\btau^a\Psi\right)^{\dagger}\vec{\nabla}\Psi
-\bigl(\vec{\nabla}\Psi\bigr)^{\dagger}(\btau^a\Psi) \Bigr]
-\Psi^{\dagger} \Bigl( \frac{g^2}{4m}\vec {W}^{a}\Bigr)\Psi
\ee
The second term is the non-Abelian analog of the London term in 
superconductivity. The main result of this approach is then to recognize that 
such second term exactly cancels the symmetry breaking term in 
Eq.\ref{totalspatialcurrent} and renders {\it zero} matter currents 
proportional to the electric field (zero spin conductivity in arbitrary space
dimension). The scenario 
is now very different from superconductivity: there, the London term is the 
only one remaining after symmetry breaking, while for the non-Abelian case, 
the London contribution gets cancelled. As discussed in references 
\cite{Tokatly,RashbaEquil}, equilibrium currents remain in relation to the 
leftover radiative contribution, cubic in the non-Abelian potential plus a 
field independent matter contribution.

\section{Abelian analogy\label{Sec:Abelian}}
In order to discuss an analogy, we will consider the simpler case of the
$U(1)$ Abelian gauge theory. Then, the Lagrangian density reads 
as~\cite{Tinkham}
\be
{\mathcal L}=i\hbar\psi^*\partial_t\psi
-\frac 1{2m}(-i\hbar{\vec {\mathcal D}}\psi)^*
(-i\hbar{\vec {\mathcal D}}\psi)-\psi^*(e\phi)\psi+\frac 1{2\mu_0}{\vec B}^2-
\frac 12\varepsilon_0{\vec E}^2\label{L-abelian}
\ee
and minimization of the action with respect
to the gauge field leads to
\be
\frac{i\hbar e}{2m}(\psi{\vec \nabla}\psi^*-\psi^*{\vec \nabla}\psi)
-\frac{e^2}m{\vec A}\psi^*\psi
-\frac 1{\mu_0}{\vec\nabla}\times{\vec B}=0
\ee
from which 
the charge current density follows
\be
{\vec j}=
\frac{e}{2m}(\psi^*(-i\hbar {\vec \nabla}\psi)
+(-i\hbar {\vec \nabla}\psi)^*\psi)
-\frac{e^2}m{\vec A}\psi^*\psi.\label{eq-jU1}
\ee
The first term 
is usually referred to as the paramagnetic term, while the second
term is responsible for diamagnetic properties of matter.
If one would have changed the Lagrangian in eq.~\ref{L-abelian}, adding a 
GSB term as
\be
{\mathcal L}\longrightarrow {\mathcal L}+\frac{e^2}{2m}{\vec A}^2\psi^*\psi,
\ee
the current would have been changed to its paramagnetic contribution only, 
{\it i.e.} without any dependence on the gauge field (no second term in the 
r.h.s. of equation~\ref{eq-jU1}). This is what occurs, as a first 
approximation, in a paramagnetic metal in a weak magnetic field.
The cancellation of this diamagnetic
term corresponds to the similar scenario of the vanishing of any spin Hall
effect in the Pauli SO case.

\section{Some implications\label{Sec:Outcomes}}
\subsection{Vanishing of the spin Hall conductivity for Rashba materials
in arbitrary dimension}
As we have mentioned, the presence of the 
gauge symmetry breaking term in the Lagrangian density exactly compensates the
``diamagnetic'' (also called diacolor) term in the spin current density,
and therefore there cannot be any spin current proportional to the electric
field~\cite{Medina}. 
This means that the spin Hall conductivity identically vanishes, and
this result is true in arbitrary dimensions.
In the particular case of a two-dimensional system with Rashba SO interaction,
Rashba has shown, using sum rule arguments, that there is no spin Hall
conductivity~\cite{Rashba}. Our conclusion is more general in the sense 
that in any dimension
we obtain a spin Hall conductivity which vanishes due to an exact cancellation
between two terms. In two dimensions, the situation is special in the sense
that both terms do in fact vanish already!

\subsection{Voltage quantization}
We now consider a two-dimensional electron gas (2DEG) is a crystal, and we
analyze the consequence of periodicity in the real space. If we consider the
transport of the spinor~\cite{WuYang} along the primitive cell of the crystal, 
we have
to form quantities like
\be\exp{\Bigl(i\hbar^{-1}\oint(\vec p+g\vec W^a\btau^a)d\vec r\Bigr)},\ee
in the absence of magnetic field.
If we restrict ourselves to transport along vectors $\vec a$ and $\vec b$
(or along $\vec b$, then along $\vec a$) in
a uniform external electric field, the commutator $[T_{\vec a},T_{\vec b}]$ 
should be considered, where $T_{\vec a}=
\exp{\Bigl(i\hbar^{-1}(\vec p+g\vec W^a\btau^a)\vec a\Bigr)}$.
This commutator can be shown to be proportional to 
$\sin\frac{e}{2mc^2}|\vec a\times\vec E| 
\sin\frac{e}{2mc^2}|\vec b\times\vec E|$, thus $[T_{\vec a},T_{\vec b}]$ 
vanishes when $\vec E$ is in the plane of the 2DEG~\cite{Medina}. 
On the other hand, 
when $\vec E$ is 
perpendicular to the plane, a quantization condition appears for the 
voltage along at least an arm of the elementary cell, say the arm $a$ 
\be E_za=p\pi mc^2/e
\ee
with $p$ an integer.
The quantity $2\pi/(2mc^2/e)$ plays the role of a quantum of voltage similar
to the flux quantum in the Aharonov-Bohm (AB) effect~\cite{Medina}. 
The analogy is nevertheless not complete, and there are important differences.
While the gauge invariant phase in the AB effect is given by 
$\oint \vec A d\vec r$ along a closed path, in the present situation 
the corresponding quantity reduces to the integral along an open path,
$\int |\vec E\times d\vec r|$. This is of course due to the fact that
the non-Abelian gauge field is given by the electric field itself, which is 
gauge invariant already (and there is no need to close the path to render
the phase gauge invariant). The present case is thus more similar
to the Aharonov-Casher effect~\cite{AharonovCasher}.

Let us mention that in a semi-conductor, using the effective mass of the 
electron instead of its bare mass, the quantum of voltage would be considerably
reduced.

\subsection{Spin interferometry}
In this section we consider an electronic Mach-Zehnder interferometer where 
electron beams can interfere and are then collected in two distinct 
detectors~\cite{MZI-exp}.
A magnetic field perpendicular to the plane of the interferometer creates
a gauge vector $\vec A$ in the whole space. The magnetic field could be limited
to a narrow area and does not ``touch'' the arms of the interferometer.
In the illustrative case below we consider that the electron mirrors and beam 
splitters, realized by gate potentials, are diagonal in spin space, i.e. 
they do not mix spin components. Consideration of the changes in the 
electron propagation direction on the spin components~\cite{MZI,FeveEtAl2002} 
does not change the qualitative scenario drawn below.

\begin{figure}
  \includegraphics[height=.3\textheight]{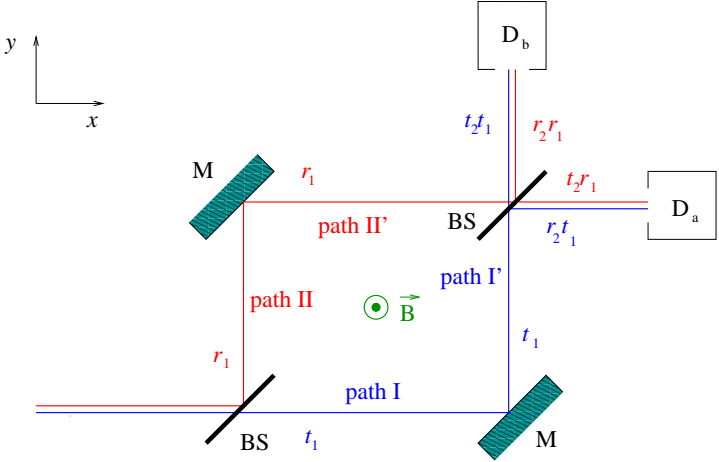}
  \caption{Mach-Zehnder electron interferometer: an
electron beam is first split into two distinct beams through a Beam Splitter
(BS), each of the beams then being reflected on a Mirror (M) before being 
collected at another BS. Two detectors (D) can measure the beams intensity in
two perpendicular directions. An external magnetic field is applied in the
perpendicular direction.}\label{FigMZ}
\end{figure}

Let us first consider the Aharonov-Bohm situation where the spin of the 
electrons is neglected, and we only discuss the case of
one detector. The electrons can follow one of the two paths called
I and II in Fig.~\ref{FigMZ}. 
The two electron paths are supposed to be essentially one-dimensional.

The wave function, transported along each path
to detector $D_a$ 
reads as
\beq
\psi_a&=&r_2\exp{\Bigl(i\hbar^{-1}\int_{I'}(\vec p+e\vec A)d\vec r}\Bigr)
	\exp{\Bigl(i\hbar^{-1}\int_{I}(\vec p+e\vec A)d\vec r}\Bigr)t_1
	\nonumber\\
	&&+t_2\exp{\Bigl(i\hbar^{-1}\int_{II'}(\vec p+e\vec A)d\vec r}\Bigr)
	\exp{\Bigl(i\hbar^{-1}\int_{II}(\vec p+e\vec A)d\vec r}\Bigr)r_1
	\Psi_0
.
\eeq

In order to calculate the probability amplitude in the detector, 
we have now to evaluate quantities like
\be
\exp\Bigl({i\hbar^{-1}\int_{I'}\Pi_y dy}\Bigr)
\exp\Bigl({i\hbar^{-1}\int_{I}\Pi_x dx}\Bigr)
\ee
with $\Pi_i=p_i+eA_i$ (take care to the non standard definition here, in order
to shorten the expression of the comparator operator~\cite{MZI}).
Using the Baker-Campbell-Hausdorff (BCH) formula and the fact that the commutator
$[\pi_x,\pi_y]$ is a $c-$number, this is equivalent to
\be
\exp\Bigl({i\hbar^{-1}\bigl(\int_{I'}\Pi_y dy+\int_{I}\Pi_x dx\bigr)
-\frac 12(i\hbar^{-1})^2\int_{I'}dy\int_I dx[\Pi_y,\Pi_x]
}\Bigr).\label{eq24}
\ee
The interferences are due to the last term involving 
$[\Pi_y,\Pi_x]=[p_y+eA_y,p_x+eA_x]=i\hbar eB$, so the last integral 
contributes to the phase by an amount $i\pi\Phi_B/\Phi_0$ with
$\Phi_0=h/|e|$ the quantum of magnetic flux (associated to the charge $e$).
Eventually, one has
\beq
\psi_a&=&\bigl(r_2t_1e^{i\pi\Phi_B/\Phi_0}+t_2r_1e^{-i\pi\Phi_B/\Phi_0}\bigr)
e^{i\alpha}\psi_0,
\eeq
where the phase $\alpha$ comes from the non-interfering contribution
in Eq.~\ref{eq24}.
This result is the standard Aharonov-Bohm effect which states that the 
interference pattern is determined by the total magnetic flux enclosed between
the two arms of the interferometer.

We now consider a variant of this problem where the spin-orbit interaction
is taken into account. 
The arms of the interferometer are supposed to be made of a
``Rashba-Dresselhaus'' active medium~\cite{MZI}.

Instead of transporting a wave function with
a phase variation induced by the coupling of the electron charge to the
Abelian gauge vector, we now consider
the transport of a Pauli spinor and its precession due to the coupling of
the electron spin to the non-Abelian gauge field.
In the case when both Rashba and Dresselhaus SO interactions are
present, the Hamiltonian
is given by the following expression~\cite{Hatano,SHChen},
\begin{equation}\label{Hamiltonian}
H= \frac{{\vec\pi}^2}{2m} + V + \alpha (\pi_x\bsigma^y-\pi_y \bsigma^x)+ 
\beta(\pi_y\bsigma^y-\pi_x\bsigma^x), 
\end{equation}
where ${\vec \pi}={\vec p}-e{\vec A}$. The non-Abelian gauge field now
takes the form 
\begin{equation}
g\vec W^a\btau^a 
= (\beta\btau^x-\alpha\btau^y)\vec {u_x}+(\alpha\btau^x-\beta\btau^y)\vec {u_y}
\end{equation}
and the spinor transport along the paths of the interferometer is defined
by the operator 
\be\Psi_{a}={\cal U}_{a}\Psi_0,\quad\ee
according to (here we disregard the fact that the spin components are mixed at each 
reflection \cite{FeveEtAl2002})
\beq
{\cal U}_a&=&r_2\exp{\Bigl(i\hbar^{-1}\int_{I'}(\vec \Pi+g\vec W^a\btau^a)d\vec r}\Bigr)
	\exp{\Bigl(i\hbar^{-1}\int_{I}(\vec \Pi+g\vec W^a\btau^a)d\vec r}\Bigr)t_1
	\nonumber\\
	&&+t_2\exp{\Bigl(i\hbar^{-1}\int_{II'}(\vec \Pi+g\vec W^a\btau^a)d\vec r}\Bigr)
	\exp{\Bigl(i\hbar^{-1}\int_{II}(\vec \Pi+g\vec W^a\btau^a)d\vec r}
\Bigr)r_1.
\eeq

Again, the evaluation of the previous expressions is made delicate due to non
commutativity of the gauge field components (and their commutator
is no longer a simple $c-$number), 
so the BCH formula is now
of no use, since it
would require the complete nested expression. We rather use properties of
the Pauli marices to decompose the exponential functions,
$\exp(\pm i \gamma \bsigma^n)=\cos \gamma\nbOne \pm i\bsigma^n\sin\gamma$.
After some algebra, we
get 
\begin{displaymath}
{\cal U}_{a}={\mathcal A}_{+}[\cos^2\Lambda-\sin^2\Lambda\sin 2\theta]
\nbOne_{2\times 2}+i(\sin\Lambda) \nbM, 
\end{displaymath}
where we have introduced the traceless matrix 
\be\nbM={\mathcal A}_{-}(\sin\Lambda\cos 2\theta)\bsigma^z-{\mathcal A}_{+}\cos\Lambda(\cos\theta+\sin\theta)(\bsigma^x-\bsigma^y),\ee 
the dimensionless variables
\begin{equation}\label{Lambda}
\Lambda=(m^*L/\hbar)\sqrt{\alpha^2+\beta^2}, 
\end{equation}
\be\theta\equiv\tan^{-1}(\beta/\alpha),\ee
and the coefficients
 \be{\mathcal A}_{\pm}=t_1t_2\pm r_1r_2 e^{2i\pi\Phi_B/\Phi_0}.\ee 
In the definition of $\Lambda$, the parameter $L$ is the length of a single arm
of the interferometer (the two arms are chosen of equal lengths).
If the parameter $\Lambda$ vanishes, the SO interaction simply disappears
and we are led to the Abelian AB situation. If $\cos\Lambda=0$, the operator
${\cal U}_a$ is diagonal in the input spinor basis, which means that a perfect
spin filtering is possible in this original basis if one of the two
eigenvalues $\lambda_\pm^a$ vanishes (in which case the corresponding
component is filtered and only the other component 
survives in detector $a$). If it is not the case, filtering may still be 
possible, but in a different basis, tilted from the original one~\cite{Hatano}.

The traceless condition simplifies the diagonalization of $\nbM$, and the eigenvalues for ${\cal U}_{a}$ are easily found to be
\begin{eqnarray}\label{evalue1}
 \lambda^{a}_{\pm}&=&{\mathcal A}_{+}[\cos^2\Lambda-\sin^2\Lambda\sin 2\theta]\mp i\sin\Lambda\times\nonumber\\
&& \sqrt{{\mathcal A}^2_{-}\sin^2\Lambda\cos^2 2\theta+2{\mathcal A}^2_{+}\cos^2\Lambda(1+\sin 2\theta)}.
 \end{eqnarray}


The filtering condition (e.g. of the $+$ component of the spinor) 
is guaranteed by the condition $\lambda_a^+=0$. It can be shown that this 
condition {\em can always be fullfiled} by a convenient choice of the magnetic
flux between the arms of the interferomter. As we have discussed above, 
an interesting feature is that, depending on the value of $\Lambda$
(remember that the Rashba amplitude can be tuned experimentally), it is
possible to achieve this condition in a non tilted basis.
When the perfect filtering condition is satisfied, the amplitude of the
non filtered component is obtained via the other eigenvalue,
$\lambda_a^-\Psi_0^-$.
In the following figures we propose illustrations in the most generic case,
i.e. the tilted basis, since this is the situation that occurs in general.
In contrast, the condition for perfect filtering in the original basis 
requires a particular relation between the Rashba and Dresselhaus amplitudes 
through the relation $\cos\Lambda=0$. 
We thus show contour plots of the magnetic flux
which allows for perfect filtering 
and for the intensity of the non
filtered component in the detector in the tilted basis,
in the plane $\alpha$, $\beta$.

\section{Conclusion}
We have presented a non Abelian gauge theory suited to deal with 
non relativistic quantum mechanics in the presence of 
various types of spin-orbit interactions. This formulation has the advantage
of a correct definition of the spin current density and is adapted to treat 
problems like topological quantization and spin filtering. This is a very
elegant way to study spintronics.

\begin{figure}
  \includegraphics[height=.3\textheight]{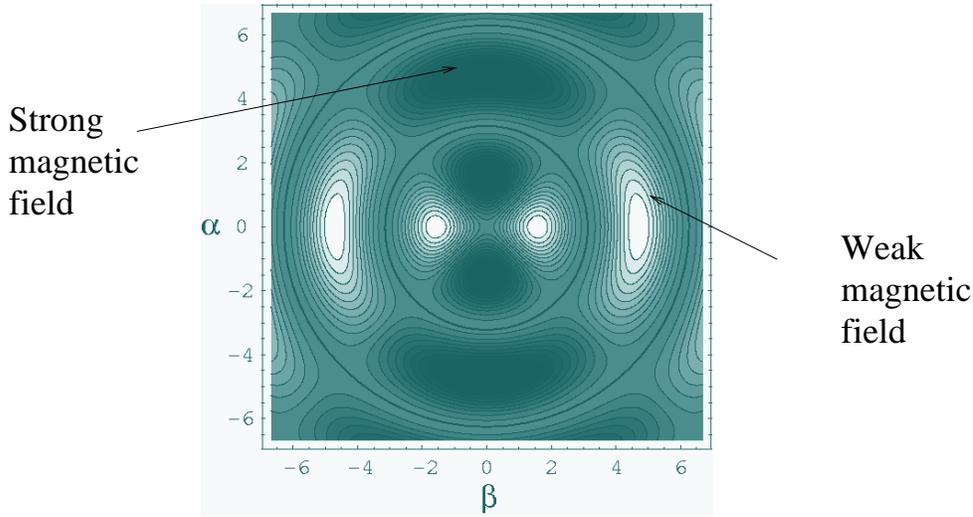}
  \caption{Perfect filtering by interference for the tilted axis. 
The plot shows a contour plot of $\sin\pi\Phi_B/\Phi_0$ in the plane 
$\alpha$, $\beta$ (in units of 
$\hbar/(m^* L)$). The darker 
regions indicate larger values for the magnetic flux needed to yield 
the condition of perfect 
filtering, from an unpolarized input. Highlighted circle and diagonal lines 
depict the zero flux solutions that yield perfect filtering.}
\label{fig2}
\end{figure}

\begin{figure}
  \includegraphics[height=.3\textheight]{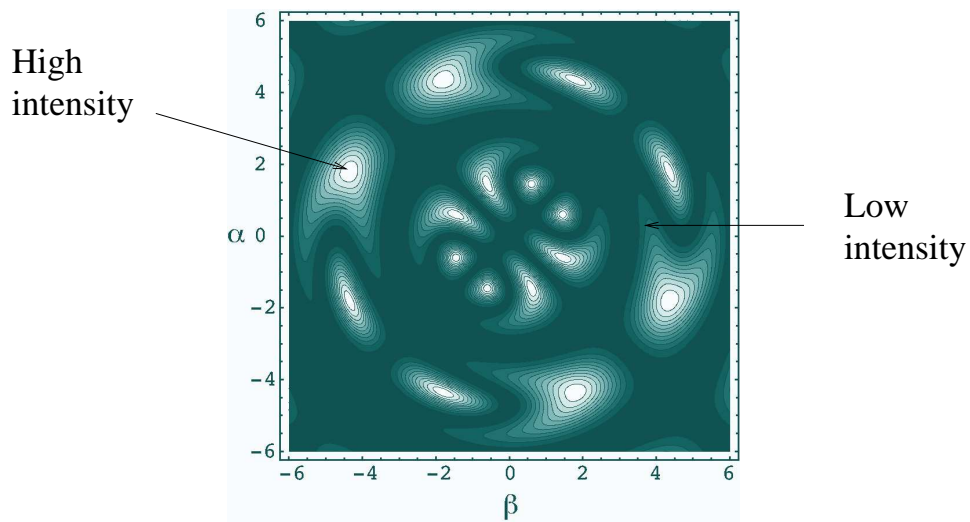}
  \caption{Perfect filtering probability for the tilted axis. The figure shows 
a contour plot (in the plane $\alpha$, $\beta$ in units of $\hbar/(m^* L)$) 
of the intensity in detector $a$, $|\Psi_a^-|^2$, assuming $|\Psi_0^-|=1$. 
Perfect spin filtering  is satisfied by the condition $\lambda_a^+=0$.
The lighter regions indicate larger values for the intensity of filtering 
for the relation between parameters depicted in Fig.\ref{fig2}. Note that 
circles 
and diagonals evident from figure \ref{fig2} correspond to zero output 
amplitude.}
\label{fig3}
\end{figure}




\end{document}